\documentclass[floatfix,aps,prd,reprint,nofootinbib,superscriptaddress,amsmath,amssymb]{revtex4-2}

\usepackage[italicdiff]{physics}
\usepackage{blindtext}
\usepackage{graphicx}
\usepackage{dcolumn}
\usepackage{hyperref}
\usepackage{bm}
\usepackage{color}

\def\eq{Eq.}
\def\eqs{Eqs.}
\def\f0{F}

\begin{document}

\title{Dynamical perturbations of black-hole punctures: effects of slicing conditions}

\author{Sean E. Li} 
\email{sli@bowdoin.edu}
\affiliation{Department of Physics and Astronomy, Bowdoin College, Brunswick, Maine 04011, USA}
\author{Thomas W. Baumgarte}
\email{tbaumgar@bowdoin.edu}
\affiliation{Department of Physics and Astronomy, Bowdoin College, Brunswick, Maine 04011, USA}
\author{Kenneth A. Dennison}
\email{kdenniso@bowdoin.edu}
\affiliation{Department of Physics and Astronomy, Bowdoin College, Brunswick, Maine 04011, USA}
\author{H. P. de Oliveira}
\email{henrique.oliveira@uerj.br}
\affiliation{Departamento de F\'isica Te\'orica, Instituto de F\'isica A. D. Tavares, Universidade do Estado do Rio de Janeiro, R. S\~ao Francisco Xavier, 524, 20550-013 Rio de Janeiro, Brazil}

\begin{abstract}
While numerous numerical relativity simulations adopt a 1+log slicing condition, shock-avoiding slicing conditions form a viable and sometimes advantageous alternative.  Despite both conditions satisfying similar equations, recent numerical experiments point to a qualitative difference in the behavior of the lapse in the vicinity of the black-hole puncture: for 1+log slicing, the lapse appears to decay approximately exponentially, while for shock-avoiding slices it performs approximately harmonic oscillation.  Motivated by this observation, we consider dynamical coordinate transformations of the Schwarzschild spacetime to describe small perturbations of static trumpet geometries analytically.  We find that the character of the resulting equations depends on the (unperturbed) mean curvature at the black-hole puncture: for 1+log slicing it is positive, predicting exponential decay in the lapse, while for shock-avoiding slices it vanishes, leading to harmonic oscillation.  In addition to identifying the value of the mean curvature as the origin of these qualitative differences, our analysis provides insight into the dynamical behavior of black-hole punctures for different slicing conditions.
\end{abstract}
\maketitle

\section{Introduction}
\label{sec:intro}

Among the most commonly used slicing conditions in numerical relativity is the {\em Bona-Mass\'o condition}
\begin{equation} \label{bona_masso}
    (\partial_t - \beta^i \partial_i) \alpha = - \alpha^2 f(\alpha) K,
\end{equation}
where $\alpha$ is the lapse function, $\beta^i$ the shift vector, $K$ the mean curvature (i.e.~the trace of the extrinsic curvature), and the {\em Bona-Mass\'o function} $f(\alpha)$ is a function of the lapse that has yet to be determined (see \cite{BonMSS95}).  The properties of the resulting slices depend, of course, on the choice for $f(\alpha)$; for $f(\alpha) = 1$, for example, the slicing condition (\ref{bona_masso}) is equivalent to the lapse condition in harmonic coordinates. 

A particularly successful choice for the Bona-Mass\'o function is
\begin{equation} \label{1+log}
    f(\alpha) = \frac{2}{\alpha},
\end{equation}
especially for simulations of black-hole spacetimes.  In the absence of a shift vector, \eq~(\ref{bona_masso}) can then be integrated to yield $\alpha = 1 + \log(\gamma)$, where $\gamma$ is the determinant of the spatial metric, which lends this slicing condition its name {\em 1+log slicing}  (see \cite{Alc08,BonPB09,BauS10,Gou12,Shi16,BauS21} for textbook discussions).  Dynamical simulations with 1+log slicing render black holes in a trumpet geometry, which, in the static limit, have been analyzed by a number of different authors \cite{HanHPBM07,HanHOBO08,Bru09}.  These studies, together with those of similar trumpet geometries (e.g., \cite{HanHBGSO07,BauN07,DenB14,BaudeO22}) have helped explain the remarkable numerical properties of these slicing conditions, especially in the context of black-hole simulations.

\begin{figure}
    \centering
    \includegraphics[width = 0.45 \textwidth]{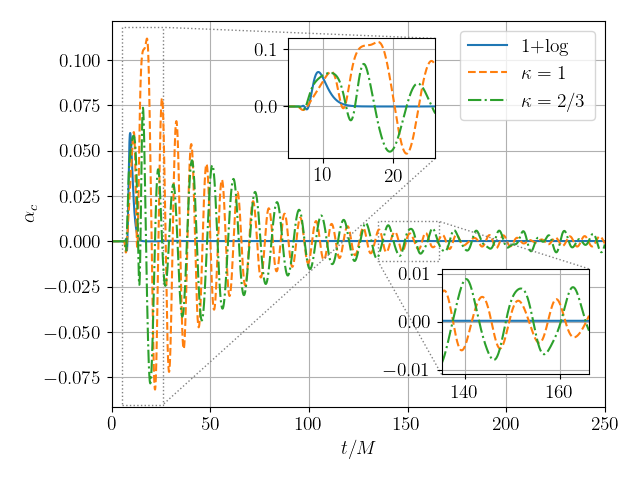}
    \caption{The lapse $\alpha$ at the black-hole puncture in the evolution of a single black hole with the Bona-Mass\'o slicing condition (\ref{bona_masso}).  All simulations start with wormhole initial data together with a ``pre-collapsed lapse", $\alpha = \psi_0^{-2}$, where $\psi_0$ is the initial conformal factor, and transition to a trumpet geometry determined by the choice of the Bona-Mass\'o function $f(\alpha)$.  Note the qualitatively different behavior of the lapse for different functions $f(\alpha)$ after the initial perturbation: for 1+log slicing with (\ref{1+log}), the lapse appears to decay approximately exponentially, while for shock-avoiding slicing conditions (\ref{shock}) it appears to perform harmonic oscillations, with a period that appears to depend on $\kappa$.  (Figure adopted from  Fig.~2 of \cite{BauH22}, to where the reader is referred for numerical details.)} 
    \label{fig:lapse}
\end{figure}

Even in the context of vacuum evolution calculations, however, 1+log slicing is known to lead to coordinate shocks in some circumstances (see \cite{Alc97,AlcM98,Alc05}).  Alcubierre \cite{Alc97,Alc03} therefore suggested an alternative {\em shock-avoiding} Bona-Mass\'o slicing condition with
\begin{equation} \label{shock}
    f(\alpha) = 1 + \frac{\kappa}{\alpha^2},
\end{equation}
where $\kappa > 0$ is a constant.  While this condition has indeed been found to avoid some coordinate pathologies that arise in 1+log slicing, it also has some very unusual properties---in particular, it allows the lapse to become negative (see the discussion in \cite{Alc03}, as well as Fig.~\ref{fig:lapse} below for an example), which may explain why it has been adopted only rarely (see, e.g., \cite{JimVA21}).  

Despite the appearance of negative values for the lapse, shock-avoiding slicing has recently been shown to perform very similarly to 1+log slicing in terms of stability and accuracy for a number of test calculations involving black holes, neutron stars, and gravitational collapse (see \cite{BauH22}).  One of these tests considered a Schwarzschild black hole initially represented on a slice of constant Schwarzschild time, i.e.~in a wormhole geometry.  These data are then evolved with the Bona-Mass\'o slicing condition (\ref{bona_masso}), which results in a coordinate transition to a trumpet geometry.  In Fig.~\ref{fig:lapse} we reproduce results from this test and show the values of the lapse at the black-hole puncture, i.e.~at the center of the isotropic coordinate system.  

Evidently, the behavior of the lapse at the black-hole puncture for 1+log versus shock-avoiding slices shows not only quantitative but also qualitative differences.  For 1+log slices the lapse appears to decay approximately exponentially after a brief dynamical period, while, for shock-avoiding slices, the lapse appears to perform harmonic oscillations.  At early times these oscillations appear to be damped, but at later times the amplitude remains approximately constant.  We also observe that the period of the oscillations is larger for a smaller value of the constant $\kappa$ in (\ref{shock}).

We caution that neither the exponential decay nor the harmonic oscillation is exact.  We also note that, because of the lack of differentiability at the center of the black hole, numerical error arising from finite-differencing across the black-hole puncture is large and prevents pointwise convergence.  Using a completely independent code based on a multi-domain spectral method (see \cite{deO22}) we found some quantitative differences resulting from the different treatment of the puncture, but the same qualitative behavior as with the finite-difference code: exponential decay, typically associated with a first-order ordinary differential equation, for 1+log slicing, versus harmonic oscillation, pointing to a second-order equation, for shock-avoiding slicing.  Since both slicing conditions are imposed by the same equation, the Bona-Mass\'o condition (\ref{bona_masso}), the origin of this qualitatively different behavior is, a priori, not clear at all.  

Our goal in this paper is to gain analytical insight into what causes these qualitative differences.  We employ a dynamical height-function approach to describe time-dependent coordinate transformations of Schwarzschild black holes, and to explore the behavior of the lapse at the black-hole puncture.  We introduce this formalism in Sec.~\ref{sec:height_function}, and review results for static slices in Sec.~\ref{sec:static}.  In Sec.~\ref{sec:perturbation} we then consider dynamical slices in the limit that they can be considered small perturbations of static slices.  At large distances from the black hole, the Bona-Mass\'o condition (\ref{bona_masso}) results in well-known wave equations for the lapse, as expected. At the black-hole puncture, however, the resulting equation depends on whether or not the (unperturbed) mean curvature $K$ vanishes at the puncture.  Typically, including for 1+log slicing, $K$ is positive at the puncture, in which case one obtains exponential damping.  Shock-avoiding slices, however, form an exception in that $K$ vanishes at the puncture, in which case one obtains harmonic oscillation.  We briefly summarize in Sec.~\ref{sec:summary}, concluding that the vanishing of $K$ at the black-hole puncture results in the qualitative differences observed.

\section{Dynamical height functions}
\label{sec:height_function} 

We start with the Schwarzschild line element in Schwarzschild coordinates,\footnote{We adopt geometrized units with $G = 1 = c$ unless noted otherwise.}
\begin{equation} \label{lineelement1}
ds^2 = - \f0 \dd{\bar{t}}^2 + \f0^{-1} \dd{R}^2 + R^2 \dd{\Omega}^2,
\end{equation}
where $R$ is the areal radius, $\f0 = \f0(R) = 1 - 2M/R$, and $M$ is the black-hole mass.\footnote{We focus on uncharged Schwarzschild black holes here, but note that our calculation generalizes to charged Reissner-Nordstr\"om black holes simply by letting $\f0 = 1 - 2M/R + Q^2 / R^2$, where $Q$ is the black-hole charge.}  We then transform to a new time coordinate\footnote{Unlike in \cite{LiBDdeO22}, where we denoted the Schwarzschild time as $t$ and the new time coordinate as $\bar{t}$, we here adopt the opposite convention in order to reduce notational clutter for dynamical slices.} $t$ by introducing a {\em height-function} $h(t,R)$ that measures how far the new time slices lift off the old time slices,
\begin{equation} \label{heightfunction}
t = \bar{t} + h(t,R)
\end{equation}
(see, e.g.,~\cite{rei73,BeiO98,MalO03,GarGH08}, as well as \cite{BauS10} for a textbook treatment).  Unlike in many previous applications, we allow the height-function to depend on time in order to study dynamical coordinate transitions.  Inserting (\ref{heightfunction}) into the line element (\ref{lineelement1}) results in 
\begin{align} \label{lineelement2}
ds^2 = & - \f0\qty(1 - \dot h)^2 dt^2 + 2 \f0 \qty(1 - \dot h) h' dt dR \nonumber \\
& + \qty(\f0^{-1} - \f0 h'^2) dR^2 + R^2 d\Omega^2,
\end{align}
where the dot denotes a partial derivative with respect to time and a prime with respect to areal radius $R$.  From (\ref{lineelement2}) we can identify the $RR$-component of the spatial metric $\gamma_{ij}$ as
\begin{subequations} \label{metric}
\begin{equation} \label{gamma1}
\gamma_{RR} = \f0^{-1} (1 - \f0^2 h'^2),
\end{equation}
the lapse function $\alpha$ as
\begin{equation} \label{alpha1}
    \alpha^2 = \frac{\f0(1 - \dot h)^2}{1 - \f0^2 h'^2},
\end{equation}
and the $R$-component of the shift vector $\beta^R$ as
\begin{equation} \label{beta1}
    \beta^R = \frac{\f0^2 (1 - \dot h) h'}{1 - \f0^2 h'^2}.
\end{equation}
\end{subequations}
Finally we compute the mean curvature from
\begin{equation} \label{K1}
    K = - \nabla_a n^a = - |g|^{-1/2} \partial_a \qty(|g|^{1/2} n^a),
\end{equation}
where $\nabla_a$ is the covariant derivative associated with the spacetime metric, $n^a$ the future-oriented normal to the spatial hypersurface, $n^a = \alpha^{-1}(1, - \beta^i)$, and $g$ the determinant of the spacetime metric, $g = - \alpha^2 \gamma_{RR} R^4 \sin^2 \theta$.   

While the height-function approach has been adopted to study the Schwarzschild spacetime in many different coordinate systems, we focus here on transformations to trumpet geometries that satisfy the Bona-Mass\'o slicing condition (\ref{bona_masso}). 

\section{Static slices}
\label{sec:static}

The construction of static trumpet geometries using a time-independent height function $h = h_0(R)$ has been discussed by a number of authors (see, e.g., \cite{HanHBGSO07,HanHPBM07,BauN07,HanHOBO08,Bru09,DenB14,BaudeO22,LiBDdeO22}), and we therefore review only some important results that are relevant for our discussion in the following sections.

For static slices, the Bona-Mass\'o condition (\ref{bona_masso}) is
\begin{equation} \label{bona_masso_static}
    \beta^i \partial_i \alpha = \alpha^2 f(\alpha) K,
\end{equation}
the expression (\ref{gamma1}) for the $RR$-component of the spatial metric remains unchanged, \eq~(\ref{alpha1}) for the lapse reduces to\footnote{We note a typo in the corresponding \eq~(7) of \cite{BaudeO22}, where the factor $\f0$, denoted $f_0$ there, should be squared in the denominator.}
\begin{subequations} \label{metric_static}
\begin{equation} \label{alpha_static}
    \alpha^2 = \frac{\f0}{1 - \f0^2 h_0'^2} = \gamma_{RR}^{-1},
\end{equation}
and \eq~(\ref{beta1}) for the shift becomes
\begin{equation} \label{beta_static}
    \beta^R = \frac{\f0^2 h_0'}{1 - \f0^2 h_0'^2} = \alpha \sqrt{\alpha^2 - \f0}.
\end{equation} 
\end{subequations}
Inserting the above expressions together with (\ref{K1}) into (\ref{bona_masso_static}) then yields an ordinary differential equation that, for many choices of the Bona-Mass\'o function $f(\alpha)$, can be integrated in closed form.  A constant of integration can be determined by imposing regularity across a singular point, making the solution unique.  For some choices of $f(\alpha)$, this solution can be expressed as an explicit function $\alpha = \alpha(R)$, but for others the solution can be written in implicit form for $\alpha$ only.  

In either case we may find the location $R_0$ of the root of the lapse, $\alpha(R_0) = 0$, which must be inside the horizon, i.e.~$R_0 < 2M$, for horizon-penetrating slices.  Defining 
\begin{equation} \label{a1}
a_1 \equiv \eval{\left( \frac{d\alpha}{dR}  \right)}_{R = R_0}
\end{equation}
we see from (\ref{alpha_static}) that $\gamma_{RR} \simeq a_1^{-2} (R - R_0)^{-2}$ close to the root of the lapse.  Assuming that $a_1$ is positive and finite, we may integrate $ds = \gamma_{RR}^{1/2} dR$ to find that the root is located at an infinite proper distance from all points $R > R_0$.  We therefore refer to this location as the {\em puncture} and note that, in its vicinity, the height function diverges according to
\begin{equation} \label{h_prime}
    h_0' \simeq - \frac{1}{\sqrt{-\f0(R_0)} \, a_1 (R - R_0)},
\end{equation}
where we have adopted a negative sign in taking a square root (note also that $\f0(R_0) < 0$ since $R_0 < 2M$).  Even for the time-dependent slices in the following sections, we will identify the puncture with a divergence of the metric component $\gamma_{RR}$, which, according to (\ref{gamma1}), coincides with a divergence of the height function $h$ at $R < 2M$.  In terms of an isotropic radius $r$, which is typically employed in numerical simulations, the puncture corresponds to the origin $r = 0$. For static slices the divergence of $h_0$ automatically coincides with a root of the lapse, but this need not be the case for time-dependent slices (see Sec.~\ref{sec:perturbation} below).  Also note that, for horizon-penetrating slices, $\alpha$ is nonzero and finite on the horizon, where $\f0 = 0$, so that (\ref{alpha_static}) indicates that the height-function necessarily has to diverge there, too.\footnote{This divergence could have been avoided by starting with a horizon-penetrating coordinate system in (\ref{lineelement1}), rather than with Schwarzschild coordinates.}  Finally we note that we can compute the static mean curvature at the puncture from
\begin{equation} \label{K_static}
    K_0(R_0) = \frac{\beta^R \partial_R \alpha}{\alpha^2 f(\alpha)} = \frac{\sqrt{- \f0(R_0)} \, a_1}{\alpha f({\alpha)}},
\end{equation}
where we have used (\ref{beta_static}) and (\ref{a1}) in the second equality.  Evidently, whether or not $K_0(R_0)$ is finite depends on the behavior of $\alpha f(\alpha)$ as $\alpha \rightarrow 0$, which we have not yet evaluated in (\ref{K_static}).

\begin{table}[t]
    \centering
    \begin{tabular}{c|c|c|c|c|c}
         $f(\alpha)$ & Ref. &  $R_0 / M$  &  $\f0(R_0)$ 
                &  $K_0(R_0) M$ & $K_0'(R_0)M^2$ \\
         \hline \hline
         $2/\alpha$ & \cite{BonMSS95} & 1.312 & $-0.524$ 
                & 0.301 & $-1.730$ \\
         $(1 - \alpha)/\alpha$ & \cite{DenB14}  & 1 & $-1$ 
                & 1 & $-3$ \\
         $1 + \kappa / \alpha^2$  & \cite{Alc97}  & 3/2 & $-1/3$ &  0 & $-8\sqrt{3}/9$ 
    \end{tabular}
    \caption{Summary of properties of static slices for three different choices of the Bona-Mass\'o function $f(\alpha)$.  For each $f(\alpha)$ we list the areal radius of the black-hole puncture $R_0$ (which, for static slices, coincides with a root of the lapse function $\alpha$), as well as the values of $\f0$, 
    the mean curvature $K_0$, and the derivative $d K / dR$ as computed from (\ref{K4}), all evaluated at the puncture.}
    \label{tab:static}
\end{table}

After this general discussion, we consider some examples for specific choices of the Bona-Mass\'o function $f(\alpha)$, and summarize the key results for these static slices in Table \ref{tab:static}.   

By far the most common choice for $f(\alpha)$ is (\ref{1+log}), which leads to {\em 1+log} slices \cite{BonMSS95}.  For 1+log slicing the integral for the lapse $\alpha$ cannot be solved for $\alpha$ directly, so that the resulting equations are usually solved numerically.  In particular, this yields $R_0 \simeq 1.312 M$ for the root of the lapse and $a_1 \simeq 0.832 M^{-1}$ for the derivative of the lapse at the root (see \cite{HanHBGSO07,HanHOBO08,Bru09}).  Finally, we use $\alpha f(\alpha) = 2$ in (\ref{K_static}) to find $K_0(R_0) = 0.301 M^{-1}$ for the mean curvature at the puncture.

\begin{figure}
    \centering
    \includegraphics[width = 0.45 \textwidth]{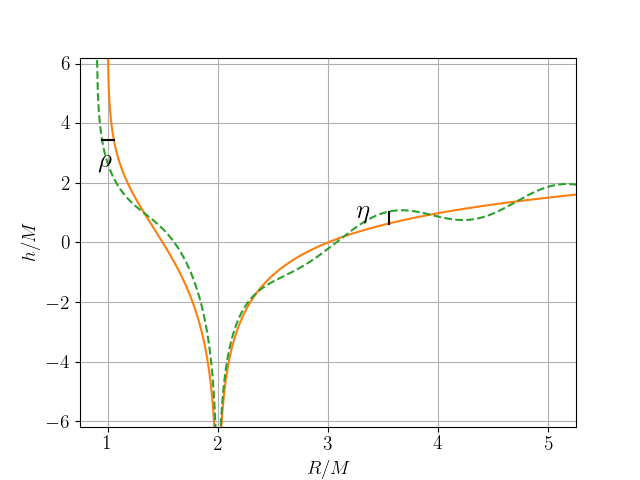}
    \caption{Graph of the static height function $h_0(R)$, \eq~(\ref{analytical_h}), for the analytical trumpet slice (solid line), together with a hypothetical perturbation (dashed line).  As discussed in Section \ref{sec:perturbation}, these perturbations can be described by changes $\eta$ in $h$ far from the puncture, and changes $\rho$ in the radius $R$ close to the puncture at $R_0 = M$.}
    \label{fig:h}
\end{figure}

As a second example we consider the choice
\begin{equation} \label{analytical_f}
    f(\alpha) = \frac{1 - \alpha}{\alpha},
\end{equation}
which results in a completely {\em analytical trumpet slicing} of the Schwarzschild spacetime (see \cite{DenB14}).  In this case the integral for the lapse can be solved explicitly, yielding
\begin{equation} \label{analytical_alpha}
    \alpha = \frac{R - M}{R}.
\end{equation}
We evidently have $R_0 = M$, from which we compute $a_1 = M^{-1}$ and, using $\alpha f(\alpha) = 1 - \alpha$, the mean curvature at the puncture, $K_0(R_0) = M^{-1}$, in agreement with \eq~(17) in \cite{DenB14}.  Because of the simplicity of the above expressions, it is also straightforward to insert (\ref{analytical_alpha}) into (\ref{alpha_static}), solve for $h_0'$ and integrate to obtain
\begin{equation} \label{analytical_h}
    h_0(R) = M \ln \frac{2 (R - 2M)^2}{(R - M) M}~~~~~~~~~~~~~~(R > M),
\end{equation}
where we have arbitrarily chosen a constant of integration so that $h_0 = 0$ at $R = 3M$.  Note that $h_0$ diverges logarithmically both at the puncture $R_0 = M$ as well as the horizon $R = 2M$, as expected from our discussion above.  In the vicinity of the puncture at $R = M$ we have $h_0' \simeq - M/(R - M)$, in agreement with (\ref{h_prime}).  We graph (\ref{analytical_h}) in Fig.~\ref{fig:h} together with a hypothetical time-dependent perturbation as considered in Sec.~\ref{sec:perturbation}.

We next consider the {\em shock-avoiding} slicing condition with $f(\alpha)$ given by (\ref{shock}).  In this case the lapse function $\alpha$ can again be expressed explicitly in terms of $R$,
\begin{equation}
    \alpha = \left( \frac{R^4 - 2MR^3 + C}{R^4 - C/\kappa} \right)^{1/2},
\end{equation}
where the constant of integration $C$ is given by $C = 3^3 M^4/2^4$ (see \cite{BaudeO22}).  The puncture is now located at $R_0 = 3 M/2$, independently of $\kappa$, and we can determine $a_1 = 2 \,(6\kappa /(3 \kappa - 1))^{1/2}/(3 M)$.  In a key difference from the other slicing conditions, however, we now observe that $\alpha f(\alpha) = \alpha + \kappa / \alpha$ diverges at the puncture, so that the mean curvature (\ref{K_static}) vanishes there, $K_0(R_0) = 0$.

\section{Dynamical slices: perturbative treatments}
\label{sec:perturbation}

We now consider dynamical slices in the limit that they may be considered linear perturbations of the static slices of Sec.~\ref{sec:static}.  Inspecting Fig.~\ref{fig:h} we note that these perturbations need to be described in different ways in different regimes.  Far from the black hole, where the slope of $h$ becomes increasingly small, the perturbed slice can be described in terms of a perturbation $\eta$ of the height function $h_0$ itself, so that $h(t,R) = h_0(R) + \eta(t,R)$.  We will briefly discuss this approach in Sec.~\ref{subsec:minkowski}, recovering well-known wave equations for the lapse function $\alpha$.  Close to the puncture, however, $h_0$ and its derivative diverge, so that changes in $h$ may also diverge.  In this region it is more natural to describe the perturbation in terms of a small shift $\rho$ in the radius $R$.  As we will show in Sec.~\ref{subsec:puncture}, this approach will yield our main result concerning the dynamical behavior of different slicing conditions at the black-hole puncture.

\subsection{Perturbations in the far limit}
\label{subsec:minkowski}

We first consider perturbations to the height function $h$ in the far limit $R \gg M$, where we assume $h_0' \rightarrow 0$.  As discussed above we describe the perturbation as 
\begin{equation} \label{perturb_far}
    h(t,R) = h_0(R) + \eta(t,R)
\end{equation}
in this regime.  Given our assumption $h_0' \ll 1$ we have $\dot h = \dot \eta$ and $h' \simeq \eta'$.  Since we also have $\f0 \simeq 1$ in this limit, we obtain, to leading order in $\eta$,
\begin{equation}
\alpha \simeq 1 - \dot \eta,~~~~~\beta^R \simeq \eta',~~~~~K \simeq \eta'' +\frac{2}{R}\eta'=\nabla^{2}\eta.
\end{equation}
Inserting these expressions into the Bona-Mass\'o condition (\ref{bona_masso}) we obtain the wave equation
\begin{equation}
    - \ddot \eta + f(1) \, \nabla^{2}\eta \simeq 0,
\end{equation}
where $f(1)$ denotes the Bona-Mass\'o function $f(\alpha)$ evaluated for $\alpha = 1$.  We may now take a time derivative of this equation to see that, in this limit, the lapse function $\alpha$ satisfies a wave equation, and that perturbations in the lapse travel at speeds $v = \sqrt{f(1)} \, c$, where we have inserted the speed of light $c$ for clarity.  For 1+log slicing with (\ref{1+log}) we have $f(1) = 2$, confirming the well-known result that gauge modes travel at a speed $v = \sqrt{2} \, c$, while for shock-avoiding slices with (\ref{shock}) gauge modes travel at a speed $v = \sqrt{1 + \kappa} \, c$ (see, e.g., \cite{Alc97,Alc03}).\footnote{Recall that the above waves describe pure gauge modes, so that a wave speed $v$ exceeding the speed of light $c$ does not violate causality.}

\subsection{Perturbations at the puncture}
\label{subsec:puncture}

We now turn to perturbations close to the puncture.  In this regime, where the static height function and its derivative diverge, a perturbative ansatz like (\ref{perturb_far}) cannot describe a perturbation like the one sketched in Fig.~\ref{fig:h}, i.e.~one that shifts the puncture to a different radius, with finite $\eta$.  Instead, we describe perturbations in the vicinity of the puncture in terms of a perturbation $\rho = \rho(t,R)$ of the radius $R$.  Specifically, we will equate the (perturbed) height function $h(t,R)$ with the static height function $h_0$ at a radius 
\begin{equation}
\bar R = R + \rho(t,R)
\end{equation}
as sketched in Fig.~\ref{fig:h}, with $\rho$ defined by
\begin{equation}
    h(t,R) = h_0\qty(\bar R) = h_0\qty(R + \rho).
\end{equation}
Derivatives of $h$ are then given by
\begin{subequations} \label{derivs_h}
\begin{align}
    \dot h(t,R) & = h_0'\qty(\bar R) \, \dot \rho(t,R) \qq{and} \\
    h'(t,R) & = h_0'\qty(\bar R) \, \left( 1 + \rho'(t,R) \right).
\end{align}
\end{subequations}
Inserting (\ref{derivs_h}) into (\ref{metric}) we obtain
\begin{subequations} \label{metric2}
\begin{equation} \label{gamma2}
    \gamma_{RR} = \f0^{-1} \left( 1 - \f0^2 h_0'^2 (1 + \rho')^2 \right)
\end{equation}
for the radial metric component,
\begin{equation} \label{alpha2}
    \alpha^2 = \frac{\f0 \qty(1 - h_0' \dot \rho)^2}{1 - \f0^2 h_0'^2 \qty(1 + \rho')^2}
\end{equation}
for the lapse, and
\begin{equation} \label{beta2}
    \beta^R = \frac{\f0^2 \qty(1 - h_0' \dot \rho) h_0' \qty(1 + \rho')}{1 - \f0^2 h_0'^2\qty(1 + \rho')^2}
\end{equation}
\end{subequations}
for the radial shift component.

We observe from (\ref{gamma2}) that $\gamma_{RR}$ diverges when $h_0'(\bar R)$ diverges.  As in the static case, we may therefore identify the puncture with a divergence of $h_0'$ at $\bar R = R_0$, except that it is now located at $R = R_0 - \rho$ (as suggested in the sketch of Fig.~\ref{fig:h}).  Evaluating the lapse (\ref{alpha2}) at the puncture, we obtain
\begin{subequations} \label{tempalphabeta}
\begin{equation} \label{alpha3a}
\alpha = (-\f0)^{-1/2} \frac{\dot  \rho}{1 + \rho'}, 
\end{equation}
while the shift (\ref{beta2}) becomes
\begin{equation} \label{beta3a}
    \beta^R = \frac{\dot  \rho}{1 + \rho'}.
\end{equation}
\end{subequations}
Defining the derivative along the normal vector $n^a$ as
\begin{equation} \label{partial_n}
    \partial_n \equiv \alpha \, n^a \partial_a = \partial_t - \beta^R \partial_R,
\end{equation}
we may rewrite \eqs~(\ref{tempalphabeta}) in the more compact form
\begin{equation} \label{alphabeta}
\alpha = (-\f0)^{-1/2} \, \partial_n \rho \qq{and} \beta^R = \partial_n \rho.
\end{equation}
Unlike in the static case (see \eq~\ref{alpha_static}), the lapse function does {\em not} necessarily vanish at the puncture for time-dependent slices, as has been observed in numerous numerical simulations (see Fig.~\ref{fig:lapse} for an example).

We next evaluate the Bona-Mass\'o condition (\ref{bona_masso}) at the puncture.  On the left-hand side we use the definition (\ref{partial_n}) together with (\ref{alphabeta}) to obtain 
\begin{equation} \label{dn_alpha}
(\partial_t - \beta^R \partial_R) \alpha = \partial_n \left((-\f0)^{-1/2} \, \partial_n \rho \right),
\end{equation}
while on the right-hand side we insert the expressions (\ref{metric2}) into (\ref{K1}) and evaluate the result at the puncture, where $h_0' \rightarrow \infty$ and $\bar R = R_0$, to obtain 
\begin{equation} \label{K3}
    K = - \frac{4 \f0 + R \f0'}{2 R (-\f0)^{1/2}}.
\end{equation}
In both (\ref{dn_alpha}) and (\ref{K3}) the function $\f0$ and its derivative are evaluated at $R = R_0 - \rho$.  For $\rho \ll R_0$ we may expand the above expressions about $\bar R = R_0$ and rewrite (\ref{K3}) as
\begin{equation} \label{K4}
    K(R) \simeq K(R_0) - \rho K'(R_0),
\end{equation}
where $K' = dK / dR$.  As expected, evaluating $K(R_0)$ from (\ref{K3}) yields the values listed in Table \ref{tab:static}, where we also list the values of $K'(R_0)$ for the different slicing conditions.  We now insert (\ref{dn_alpha}) and (\ref{K4}) into the Bona-Mass\'o condition (\ref{bona_masso}) and obtain 
\begin{equation} \label{bona_masso2}
    \partial_n \left((-\f0)^{-1/2} \, \partial_n \rho \right) = - \alpha^2 f(\alpha) \left[ K(R_0) - \rho K'(R_0) \right],
\end{equation}
where we have not yet evaluated the term $\alpha^2 f(\alpha)$.   Remarkably, all spatial derivatives of $\rho$ other than those contained in the operators $\partial_n$ on the left-hand side disappear in the limit $h_0' \rightarrow \infty$, resulting in an {\em ordinary} differential equation for $\rho$ at the puncture.  As we will explore in the next two subsections, even the qualitative behavior of solutions to this equation depends on the choice of $f(\alpha)$ and hence $K(R_0)$, because it determines whether (\ref{bona_masso2}) acts as a first- or second-order equation.

\subsubsection{1+log slicing}
\label{subsubsec:1+log}

For almost all slicing conditions, the leading-order mean curvature term $K(R_0)$ on the right-hand side of (\ref{bona_masso2}) is nonzero.  One such slicing condition is 1+log slicing with $f(\alpha) = 2 / \alpha$ (see \ref{1+log}), for which (\ref{bona_masso2}) becomes
\begin{widetext}
\begin{align} \label{bm_1+log}
    \partial_n \left((-\f0)^{-1/2} \, \partial_n \rho \right) = 2 (-\f0)^{-1/2} \, \partial_n \rho  \left[ K(R_0) - \rho K'(R_0) \right]  
\end{align}
\end{widetext}
after inserting (\ref{alphabeta}) for $\alpha$ on the right-hand side.  We now observe that, to leading order in $\rho$, the term $\rho K'(R_0)$ on the right-hand side disappears and with it the only appearance of $\rho$ itself (rather than its derivatives).
To linear order in $\rho$, we may therefore replace the term  $(-\f0)^{-1/2} \, \partial_n \rho$ with $\alpha$ to obtain a first-order equation for the lapse alone,
\begin{equation}
    \partial_n \alpha = - 2 \alpha K(R_0).
\end{equation}
This equation is identical to our starting point (\ref{bona_masso}), of course, except that now, in the linear limit, we assume the mean curvature $K$ to be given by a positive and constant value.  In this case we may integrate to obtain
\begin{equation} \label{lapse_1+log}
\alpha = C e^{-2 K(R_0) t}
\end{equation}
where $C$ is a constant of integration, demonstrating that, to linear order, we should expect the lapse function at the puncture to decay exponentially for 1+log slicing.

As one might expect from the discussion in Sec.~\ref{sec:intro}, a quantitative comparison of (\ref{lapse_1+log}) with the numerical data of Fig.~\ref{fig:lapse} shows some differences.  During the time around $10 M \lesssim t \lesssim 15 M$, when Fig.~\ref{fig:lapse} suggests an approximately exponential decay, the numerical data fall off more rapidly than predicted by (\ref{lapse_1+log}).  A rough fit to the numerical data suggests a time constant $\tau$ that is smaller than $(2 K(R_0))^{-1}$ by about a factor of two.  However, rather than being constant, $K$ also changes rapidly during this period, as it transitions from its initial value of zero to the equilibrium value of $K(R_0) \simeq 0.301$, indicating that nonlinear terms are still important during this time.  At later times numerical error becomes important; in particular, the lapse settles down to a small but nonzero value (that depends on the numerical resolution), so that exponential decay can no longer be observed.  

\subsubsection{Shock-avoiding slices}
\label{subsubsec:shockavoiding}

For shock-avoiding slices the unperturbed puncture is located at $R_0 = 3 M/2$ so that the mean curvature $K(R_0)$ on the right-hand side of (\ref{bona_masso2}) vanishes.  Inserting $f(\alpha) = 1 + \kappa / \alpha^2$ (see \ref{shock}) into (\ref{bona_masso2}) we now obtain
\begin{equation} \label{bm_sa}
    \partial_n \left((-\f0)^{-1/2} \, \partial_n \rho \right) = \kappa K'(R_0) \rho,
\end{equation}
where we have already neglected a term quadratic in $\alpha$ on the right-hand side.  In contrast to 1+log slicing, the term proportional to $\rho$ now dominates the right-hand side, so we cannot rewrite this second-order equation as a first-order equation for $\alpha$.  We instead expand to linear order in $\rho$ to obtain the harmonic-oscillator equation
\begin{equation} \label{bm_sa2}
    \partial^2_n \rho = - \omega^2 \rho,
\end{equation}
with the angular frequency $\omega$ given by
\begin{equation}
    \omega^2 = - (-\f0)^{1/2} \kappa K'(R_0) = \frac{8 \kappa}{9 M^2}.
\end{equation}
Accordingly, $\rho$, and hence $\alpha$, performs harmonic oscillations with period
\begin{equation} \label{period}
    P = \frac{3 \pi M}{\sqrt{2 \kappa}}. 
\end{equation}
Note that we have assumed $\kappa > 0$ in the above arguments, in accordance with our original assumption in (\ref{shock}) (see also \cite{Alc97}).

As in Sec.~\ref{subsubsec:1+log} we attempt a quantitative comparison with the numerical data with some caution.  Measuring the period of the oscillations observed for the shock-avoiding slices around $130 M \lesssim t \lesssim 170 M$, we find $P_1 \simeq 8 M$ for $\kappa = 1$ and $P_{2/3} \simeq 11 M$ for $\kappa = 2/3$ (even though the latter, in particular, shows some variation).  Evaluating (\ref{period}), on the other hand, we obtain $P_1 \simeq 6.7 M$ and $P_{2/3} \simeq 8.2 M$.  While we again do not find accurate quantitative agreement, we see that our analysis does explain the origin of the observed harmonic oscillation and correctly predicts that the period increases with decreasing $\kappa$.  

We suspect that nonlinear terms cause the damping of the oscillations at early times, as seen in Fig.~\ref{fig:lapse}.  Once the amplitude is sufficiently small, however, the oscillations should be governed by (\ref{bm_sa2}), which does not have a damping term. Accordingly, one would expect these oscillations to persist at a small amplitude, which is consistent with the numerical results.

\section{Summary}
\label{sec:summary}

Motivated by recent numerical experiments with shock-avoiding slicing conditions as alternatives to 1+log slicing, we explore the origins of a qualitative difference observed in these simulations: while, for the latter, the lapse function at the black-hole puncture appears to decay in an approximately exponential fashion, the former leads to approximately harmonic oscillations in the lapse.  We apply a dynamical height-function approach to Schwarzschild black holes to describe time-dependent coordinate transitions, impose the Bona-Mass\'o condition (\ref{bona_masso}) with different choices for the function $f(\alpha)$, evaluate the resulting equation at the black-hole puncture, and finally consider small perturbations of a background trumpet solution.  

Describing these perturbations in terms of the displacement $\rho$  of the location of the puncture, the Bona-Mass\'o equation becomes a second-order equation for $\rho$ (see \ref{bona_masso2}).  The key difference between 1+log slices and shock-avoiding slices then arises from the behavior of the (unperturbed) mean curvature $K(R_0)$ at the location of the puncture.  For 1+log slicing, $K(R_0)$ takes a nonzero, positive value, in which case $\rho$ itself drops out of the equation to linear order, resulting in a {\em first-order} equation for $\partial_n \rho$ that governs exponential decay.  For shock-avoiding slices, on the other hand, $K(R_0)$ vanishes, and the right-hand side of (\ref{bona_masso2}) ends up being dominated by $\rho$ at linear order.  The equation therefore remains a {\em second-order} equation for $\rho$, resulting in harmonic oscillation.  We further observe that the period of the oscillations depends on the constant $\kappa$ in (\ref{shock}), with larger $\kappa$ resulting in a shorter period.

While a quantitative comparison of our analytical results with the numerical findings of \cite{BauH22} shows some discrepancies as discussed in Sect.~\ref{subsec:puncture}, we believe that these can be explained in terms of nonlinear effects and numerical error resulting from the lack of differentiability at the black-hole puncture.  Despite these discrepancies, our findings provide analytical insight into the dynamical behavior of the lapse at the black-hole puncture, point to the origin of qualitative differences between different slicing conditions, and predict the dependence of decay constants and oscillation periods on the given parameters. 

\begin{acknowledgements}

It is a pleasure to thank David Hilditch for comments on a previous version of this manuscript.
SEL acknowledges support through an undergraduate research fellowship at Bowdoin College. 
This work was supported in part by National Science Foundation (NSF) grant PHY-2010394 to Bowdoin College and the Coordena\c c\~ao de Aperfei\c coamento de Pessoal de N\'ivel Superior - Brasil (CAPES) - Finance Code 001.
\end{acknowledgements}


%

\end{document}